# Extremely large magnetoresistance in topological semimetal candidate pyrite PtBi$_2$

Wenshuai Gao[1,2#], Ningning Hao[1#], Fa-Wei Zheng[3#], Wei Ning[1*], Min Wu[1,2], Xiangde Zhu[1†], Guolin Zheng[1,2], Jinglei Zhang[1], Jianwei Lu[1,2], Hongwei Zhang[1,2], Chuanying Xi[1], Jiyong Yang[1], Haifeng Du[1], Ping Zhang[3,4] Yuheng Zhang[1,5], Mingliang Tian[1,5‡]

[1]*High Magnetic Field Laboratory, Chinese Academy of Sciences, Hefei 230031, Anhui, China;*

[2]*Department of physics, University of Science and Technology of China. Hefei 230026, China;*

[3]*Institute of Applied Physics and Computational Mathematics, Beijing 100088, China*

[4]*Beijing Computational Science Research Center, Beijing 100193, China*

[5]*Collaborative Innovation Center of Advanced Microstructures, Nanjing University, Nanjing 210093, China*

[#] Those authors contribute equally to this work

[*]   ningwei@hmfl.ac.cn
[†]   xdzhu@hmfl.ac.cn
[‡]   tianml@hmfl.ac.cn

# Abstract

While pyrite-type PtBi$_2$ with face-centered cubic structure has been predicted to be a three-dimensional (3D) Dirac semimetal, experimental study on its physical properties remains absent. Here we report the angular-dependent magnetoresistance (MR) measurements of PtBi$_2$ single-crystal under high magnetic fields. We observed extreme large unsaturated magnetoresistance (XMR) up to 11.2 million percent at $T$ = 1.8 K in a magnetic field of 33 T, which surpasses the previously reported Dirac materials, such as WTe$_2$, LaSb and NbP. The crystals exhibit an ultrahigh mobility and significant Shubnikov-de Hass (SdH) quantum oscillations with nontrivial Berry's phase. Analysis of Hall resistivity indicates that the XMR can be ascribed to the nearly compensated electron and hole. Our experimental results associated with the *ab initio* calculations suggest that pyrite PtBi$_2$ is a topological semimetal candidate which might provide a platform for exploring topological materials with XMR in noble metal alloys.

Recent discovered three-dimensional (3D) Dirac and Weyl semimetals represent a new state of topological quantum matter. These materials are characterized with a linear energy dispersion in bulk and can be viewed as a "3D graphene"[1,2,3,4,5]. The Dirac point can be viewed as a pair of Weyl points coinciding in momentum space, protected by time reversal, space inversion as well as crystalline point-group symmetries[3,4,5,6]. Dirac semimetals can be evolved into Weyl semimetals when either time-reversal symmetry or space inversion symmetry is broken. Several materials have been confirmed to be Dirac semimetals, such as $Cd_3As_2$[7,8,9,10,11] and $Na_3Bi$[12,13]. Besides, $WTe_2$[14], TaAs[15,16,17], NbAs[18], TaP[19] and NbP[20] have also been experimentally suggested to be the Weyl semimetals. The unique band structures of these materials result in peculiar quantum phenomena, such as ultra-high carrier mobility[21,22,23] and very large non-saturating magnetoresistance (MR)[21,22]. For examples, in $WTe_2$[14], a MR of 13 million percent was observed at 0.53 K and $\mu_0H = 60\ T$. In $Cd_3As_2$, the carrier mobility is up to $9 \times 10^6\ cm^2\ V^{-1} s^{-1}$ at 5 K[11]. The MR of NbP can reach 8.1 million at high field 60 T and 1.5 K[21]. These appealing transport properties have generated immense interests in both the condensed matter physics and potential application.

Discovering new Dirac materials with novel properties has become an important front in condensed matter and material sciences. Pyrite-type $PtBi_2$ has been theoretically predicted to be a new 3D Dirac semimetal (DSM) by Gibson *et al.*[24], it has a face-centered cubic structure with a broken inversion symmetry. The different $C_3$ rotation eigenvalues result in band crossing and consequently a 3D Dirac point along the line Γ-R, thus interesting physics or phenomena are expected in this noble alloy. Motivated by this prediction, we grew high-quality pyrite-type $PtBi_2$ single crystals and carried out detailed magnetotransport study. We found that the temperature-dependent resistivity at various magnetic fields presents clearly the magnetic field-induced turn-on and plateau behavior at low temperatures. Under magnetic field up to 33 T, we observed extreme large MR (XMR) up to 11.2 million at 1.8 K without any sign of saturation, which is higher than WTe$_2$ and any other similar Dirac materials reported before[14,25,26]. Significant Shubnikov–de Hass (SdH) quantum oscillations develop at low temperatures which enables us to reconstruct the Fermi surface topology and provides insight on the nature of electronic structures in PtBi$_2$. Analysis of Hall resistivity based on two-band model indicates that the XMR can be ascribed to the nearly compensated electron and hole with high mobility at low temperatures. Our experimental results associated with the *ab initio* calculation suggest that pyrite PtBi$_2$ is a topological semimetal candidate with unique band structure that different from the previous topological semimetals.

The cubic system PtBi$_2$ crystallizes in a pyrite-type structure with a space group of $Pa\bar{3}$. The *a, b, c* axis are perpendicular to each other and the lattice parameters are $a = b = c = 6.702$Å, as shown in Fig. 1(a). The samples were prepared by the flux method and the detailed growing processes will be published elsewhere. The inset of Fig. 1(b) is the morphology of a crystal with (111) facet. Figure 1(b) shows the x-ray diffraction (XRD) oriented with the scattering vector perpendicular to the (111) plane. Magnetotransport measurements were carried out in the (111) plane using both 16 T physical property measurement system (PPMS, Quantum Design Inc) and the ~35 T dc-resistive magnet at the China High Magnetic Field Laboratory (CHMFL) in Hefei.

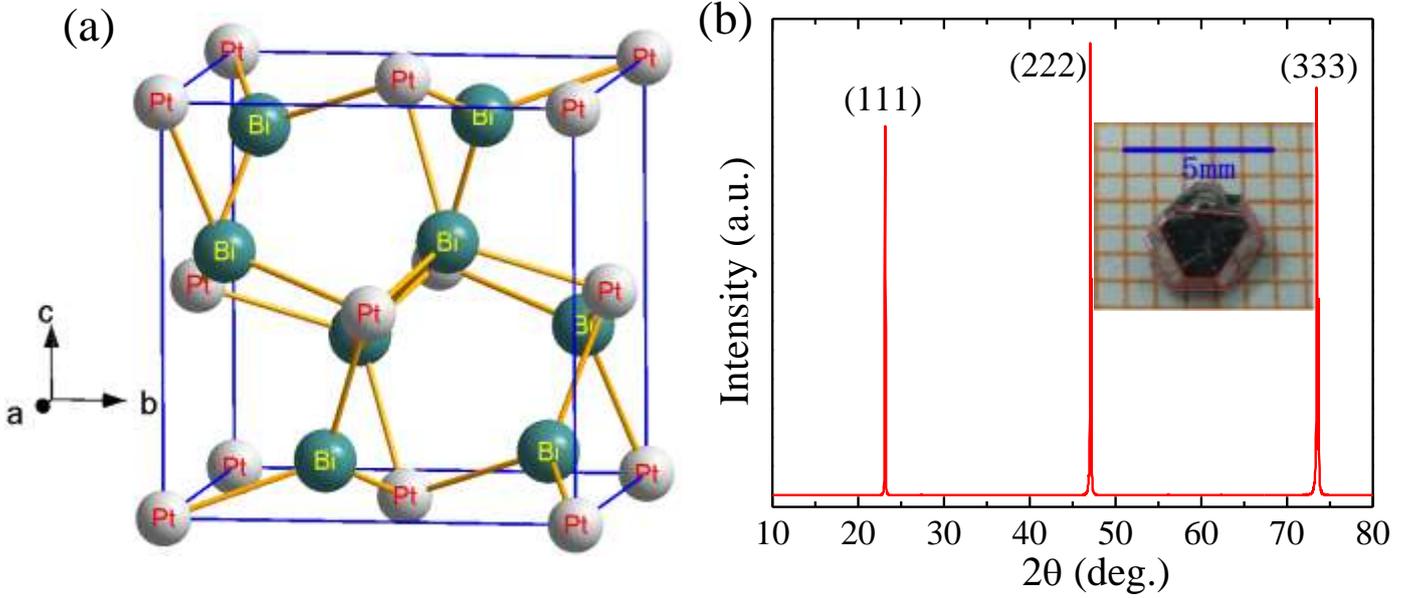

**FIG 1.** (a) The crystal structure of pyrite-type PtBi$_2$ with cubic symmetry and space group of $Pa\bar{3}$. Gray and cyan balls represent Pt atom and Bi atom, respectively. (b) The X-ray diffraction pattern of a pyrite-type PtBi$_2$ single crystal. Inset: photo of a typical PtBi$_2$ single crystal.

Figure 2(a) describes the temperature dependence of the longitudinal resistivity $\rho_{xx}$ of a sample from 2 K to 300 K at zero magnetic field. The resistivity exhibits highly metallic behavior with $\rho_{xx}(300K) = 40 \, \mu\Omega \cdot cm$ and $\rho_{xx}(2K) = 24 \, n\Omega \cdot cm$, leading to an extremely large residual resistance ratio (*RRR*), $R(300K)/R(2K) = 1667$. This result indicates that the PtBi$_2$ crystal studied here is high quality. The zero-field resistivity curve below 20 K can be well fitted by formula $\rho = \rho_0 + aT^2$, as shown in the inset of Fig.2(a), where the residual resistivity $\rho_0$ is about $18 \, n\Omega \, cm$, which is 100 times lower than WTe$_2$[14], 35 times lower than NbP[21], and comparable to that in highest-quality Cd$_3$As$_2$ crystals[27]. When a small magnetic field of 0.3 T is applied perpendicular to the (111) plane, the *R-T* curve displays a metal-insulator transition (MIT) driven by magnetic field (Fig. 2(b)) and develop a resistivity plateau at low temperature. The insulating behavior is enhanced significantly as the magnetic field increases. Such a "turn-on" and plateau behaviors at low temperatures in PtBi$_2$ is very similar to those reported in other semimetals, such as WTe$_2$[14,28], NbP[21], and LaSb[26]. To obtain the magnetic field dependence of MIT turn-on temperatures $T_m$ and resistivity plateau characteristic temperature $T_i$, we plot the $\partial\rho/\partial T$ as a function of temperature, as shown in Fig. 2(c). The characteristic temperature $T_i$, *i.e.*, the minimum of $\partial\rho/\partial T$ curves, is almost unchanged as the magnet filed increases. On the contrary, the $T_m$, which is from $\partial\rho/\partial T = 0$, increases significantly as the magnetic field increase. Such tendency is shown in the inset of Fig.2(c). Figure 2(d) is the inverse temperature dependence of $ln\Delta\rho_{xx}$. By fitting the linear region between $1/T_m < 1/T < 1/T_i$ (red line) using $\rho(T) = \rho_0 + exp\,(E_g/K_BT)$, we obtain the activation energy gap $E_g$, which is shown in the inset of Fig. 2(d) as a function of the field. The $E_g$ value increases monotonously as the magnetic field increases and reaches about 5.6 meV at $B$ = 16 T.

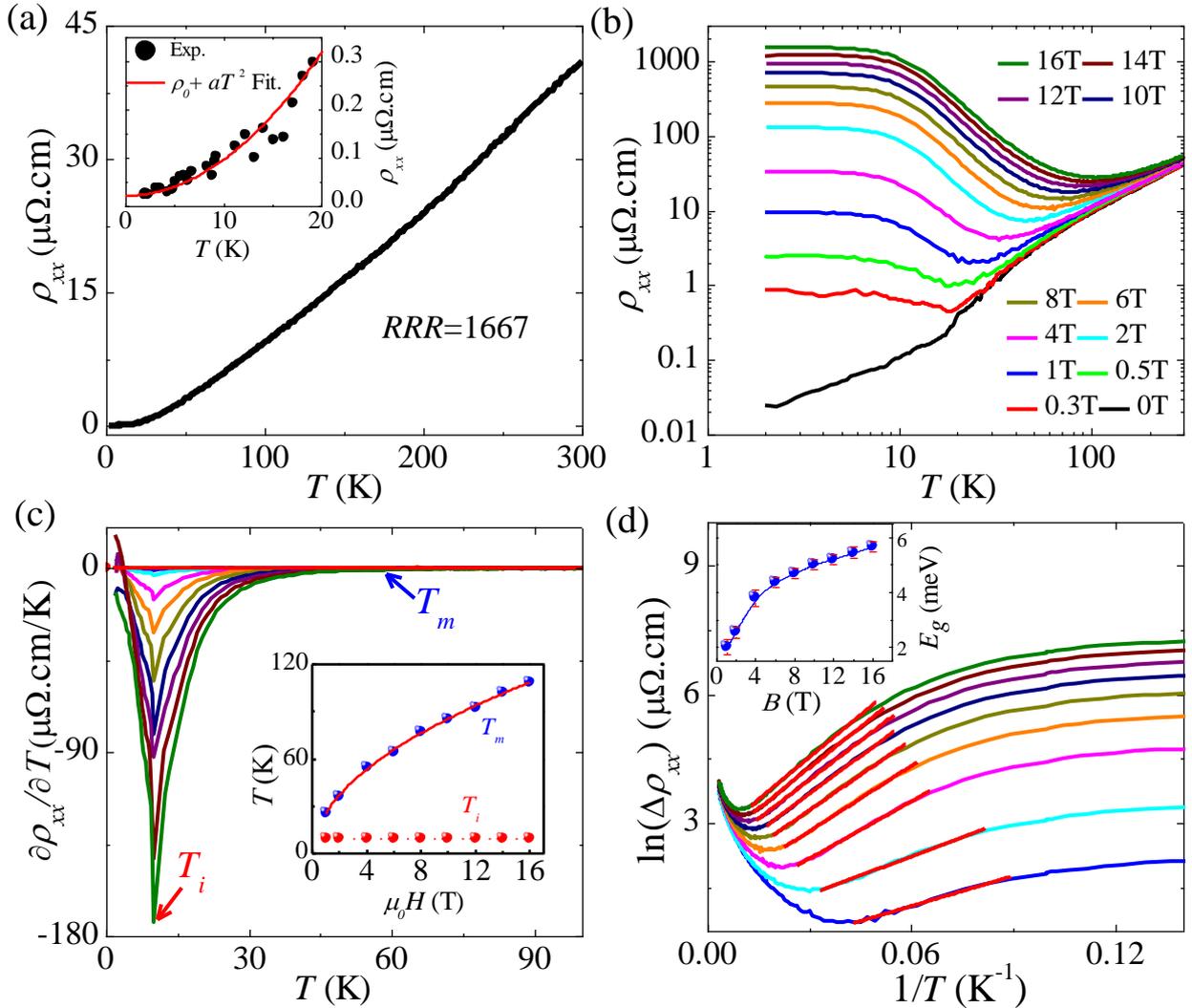

**FIG 2.** (a) Temperature dependence of the longitudinal resistivity $\rho_{xx}$ from 2 K to 300 K at zero magnetic field. Inset: The zero-field resistivity curve below 20 K can be well fitted by formula $\rho = \rho_0 + aT^2$.

(b)Temperature dependence of resistivity at various magnetic fields applied perpendicular to the (111) plane and current. (c) The $\partial\rho / \partial T$ as a function of temperature. Inset: Magnetic field dependence of characteristic temperature $T_m$ and $T_i$. (d) The inverse temperature dependence of $\ln\rho_{xx}$, the red lines are the fitting results by $\rho(T) = \rho(0) + exp\ (E_g/K_BT))$ in $1/T_m < 1/T < 1/T_i$ region. Inset: Magnetic field dependence of insulating gap value $E_g$.

Figure 3(a) shows the MR under different temperatures with the field perpendicular to high symmetry (111) plane. An extremely large MR, $[(R(B)-R(0))/R(0)] \times 100\% \sim 1.12 \times 10^7$ % is observed under 33 T at 1.8 K, which is accompanied with clear Shubnikov-de Hass (SdH) quantum oscillations. This huge MR value is comparable to those in topological semimetals WTe$_2$, LaSb and NbP, where the MR are respectively about $1.3 \times 10^7$ % under 60 T at 0.53 K[14], $\sim 1 \times 10^6$ % at 9T[26], and $\sim 8.1 \times 10^6$ % under 62 T at 1.5 K[21]. Such a large MR makes PtBi$_2$ successively into the category of the largest MR materials. Considering that the MR of PtSn$_4$[29], the only noble metal alloy showing XMR previously, is just about $5 \times 10^5$ % at 1.8 K and 14 T, PtBi$_2$ is definitely the first noble metal alloy with a MR on the scale of more than a million. Furthermore, the MR in PtBi$_2$ does not reach a saturation up to 33 T. In addition to the large magnitude, following a conventional quadratic dependence in low fields, a linear MR is also observed in our PtBi$_2$ single crystals from $B$=14 T up to 33 T. Such a linear field dependence for MR is also widely observed in these Dirac/Weyl semimetals. We have measured another sample with lower quality which shows $RRR$=524 and residual resistance $\rho_0$= 77 nΩ. cm. The XMR of this sample is 3.3 million at 33 T with a saturation trend at high field range (not shown), indicating that the crystal quality is very crucial in boosting XMR.

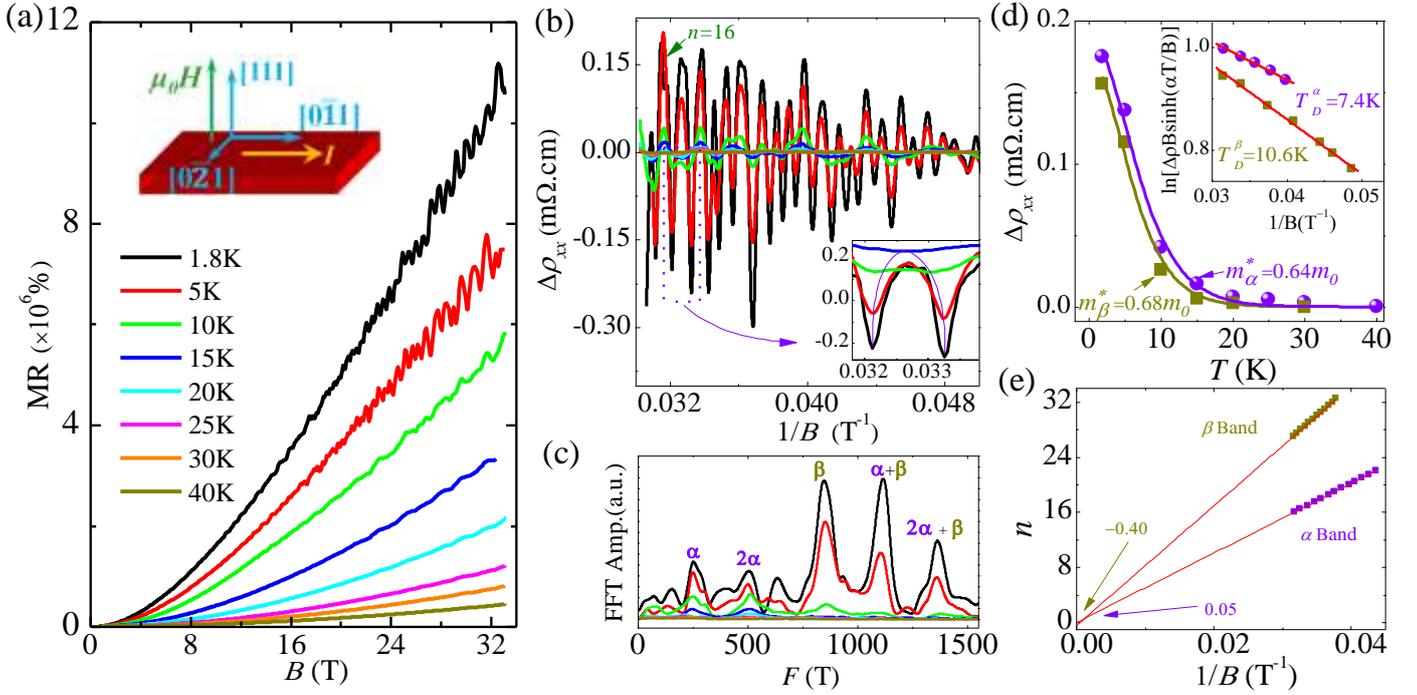

**FIG 3.** (a) MR of PtBi$_2$ under different temperatures with the field applied perpendicular to high symmetry (111) plane and current. (b) The SdH oscillation $\Delta\rho_{xx}$ plotted as a function of inverse field at different temperatures. The inset shows the Zeeman splitting of an oscillation peak at high field. Data at different temperatures has been shifted for clarity. (c) FFT spectra of SdH oscillations at different temperatures. (d) The amplitudes of the oscillatory component at various temperatures. The solid lines are fitting results of the Lifshitz-Kosevich formula. Inset: Dingle plot of the SdH oscillation associated with α and β band. (e) The Landau level indices extracted from the SdH oscillation plotted as a function of inverse field.

In order to obtain more information on the electronic structure of the PtBi$_2$, we have investigated the quantum oscillations at different temperatures with the field perpendicular to the high symmetry (111) plane. Figure 3(b) shows the SdH oscillation spectra of $\rho_{xx}$ in (111) plane after subtracting a third-order polynomial background. The oscillation amplitudes decrease gradually with increasing temperature. As shown in inset of the Fig.3 (b), the Fast Fourier Transform (FFT) spectra of SdH oscillations reveal two principal frequencies $F_\alpha$ = 250 T and $F_\beta$ = 850 T. The $F_\beta$ peak emerges at about 10 K and increases rapidly upon cooling. According to the Onsager relation $F = (\hbar/2\pi e)A_F$, the cross section area of Fermi surface $A_F$ is determined to be 0.024Å$^{-2}$ and 0.081Å$^{-2}$ for α band and β band, respectively, indicating a complicated Fermi surface. Additionally, we observed peak splitting behavior in the high field oscillations above $B = 16$ T at $T = 1.8$ K. It can be clearly seen from the inset of Fig.3b which shows a peak splitting between 0.032 T$^{-1}$ to 0.036 T$^{-1}$. With increasing temperature above 15 K, the splitting gradually merges to single peak, implying an origin of Zeeman effect. Unfortunately, we are unable to separate the $g$ factors through the Zeeman splitting since we are unable to separate electrons from different Fermi surface in SdH oscillations.

We have further estimated the effective masses of electrons as shown in Fig. 3(c), by fitting the temperature dependence of oscillation amplitudes using the Lifshitz-Kosevich (LK) formula:

$$\frac{\Delta\rho(T,B)}{\rho(B=0)} \propto \frac{2\pi^2 k_B T/\hbar\omega_c}{\sinh[2\pi^2 k_B T/\hbar\omega_c]} exp\left(-2\pi^2 k_B T_D/\hbar\omega_c\right) \quad (1),$$

where $k_B$ is the Boltzmann constant, $\hbar$ is the Planck's constant and $\omega_c = eB/m^*$ is the cyclotron frequency, with $m^*$ the effective cyclotron mass at the Fermi energy. We get the effective cyclotron mass $m_\alpha^* = 0.64 \pm 0.012\ m_0$, and $m_\beta^* = 0.68 \pm 0.014\ m_0$, respectively, with $m_0$ is the bare electron mass. Correspondingly, the Fermi velocity $v_F = \hbar k_F/m^*$, is ~$1.58 \times 10^5$ m/s for α band and ~$2.76 \times 10^5$ m/s for $\beta$ band, thus the Fermi energy $E_F = m^* v_F^2$ is $E_F^\alpha$ ~ 90.7 meV and $E_F^\beta$ ~ 292.4 meV. We have also analysis the quantum mobility, as shown in the inset in Fig. 3(c). In equation (1), the $T_D$ is the Dingle temperature which is related to the quantum scattering lifetime $\tau_Q$ by $T_D = \hbar/2\pi k_B \tau_Q$. The field dependence of the amplitude of quantum oscillations at fixed temperatures gives access to the Dingle temperature $T_D^\alpha = 7.4$ K and $T_D^\beta = 10.6$ K. The corresponding quantum scattering lifetimes are $\tau_Q^\alpha = 1.6(5) \times 10^{-13} s$ and $\tau_Q^\beta = 1.1(5) \times 10^{-13}$. The mobilities are estimated by $\mu_Q = \frac{e\tau_Q}{m^*}$ are $\mu_Q^\alpha = 453.4\ cm^2 V^{-1} s^{-1}$ and $\mu_Q^\beta = 298.7\ cm^2 V^{-1} s^{-1}$, respectively.

The nature of Dirac electrons participating in quantum oscillations can be revealed from the quantitative analysis of SdH oscillations. The nontrivial Berry phase is generally considered to be the key evidence for Dirac fermions and has been observed in other Dirac materials[30,31]. According to the Lifshitz-Onsager quantization rule[32], the oscillations can be described by $\Delta\rho_{xx} \propto \cos\left[2\pi\left(\frac{F}{B}+\phi\right)\right]$, where $F$ is the oscillations frequency and $\phi$ is the phase shift. To extract the phase shift, a plot of the Landau index $n$ vs $1/B$ then extrapolates to the phase shift on the $n$ axis. For Dirac system with linear dispersion carries an extra π Berry phase, leading to phase shifts with $\phi$ =0 in 2D system and ±1/8 in the 3D case. While for Dirac semimetals with time-reversal symmetry (e.g. $PtBi_2$), an anomalous phase shift leads the value of phase shift $\phi$ changed from ±1/8 to be ±5/8.[33] Considering $PtBi_2$ is a multi-band system and $\rho_{xx} \gg \rho_{xy}$, here the peak of the SdH oscillations should be defined as integer indices and the valley as half indices[34]. As shown in Fig. 4(e), the linear fitting of $n$ vs. $1/B$ gives the intercept value 0.05 ± 0.01 for α band, which is very close to zero and suggests a topological trivial Berry phase of α band, and -0.40±0.02 for β band (Fig. 3(e)], which is close to -5/8 and indicates a non-trivial Berry phase of β band.

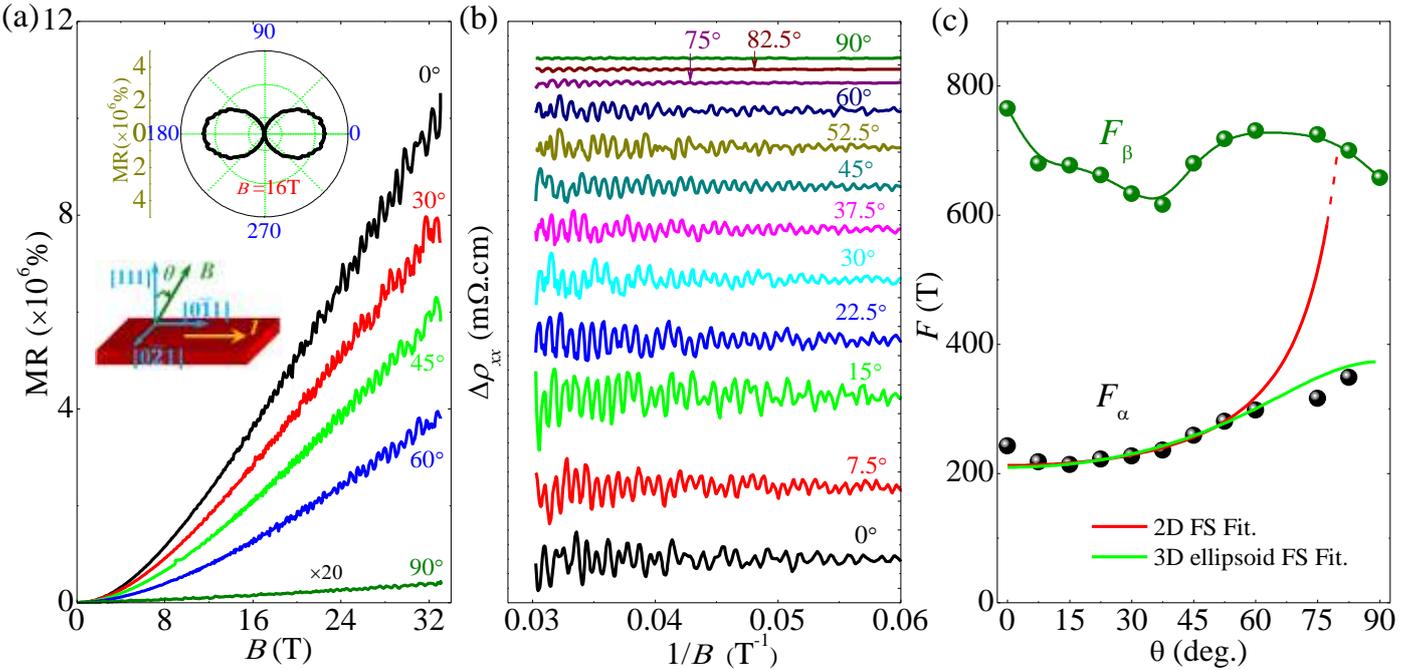

**FIG 4.** (a) Magnetic field dependent of MR under different angle, the magnetic field rotated from [111] direction to the [0$\bar{1}$1] direction parallel to the current in (111) plane. Inset shows the polar plot for the MR

under 16 T and 2 K. (b) SdH patterns at various angles plotted as a function of inverse field. (c) The angular dependence of the oscillation frequency derived from FFT analysis from (b). The solid lines are the fits to the Fermi surfaces using 2D Fermi surface and 3D ellipsoid Fermi surface model.

Figure 4(a) shows the angular-dependent MR with tilting the magnetic field from [111] direction ($\theta = 0°$) to the $[0\bar{1}1]$ direction ($\theta = 90°$) parallel to the applied current in the (111) plane. The MR decreases dramatically from $1.12 \times 10^7$ % to $2 \times 10^4$ % at $B = 33$ T, indicating a significant anisotropic MR with amplitude ratio up to $5.6 \times 10^4$ %. Such large angular-dependent MR behavior has been observed in $WTe_2$[14] and $Cd_3As_2$[21] and suggests potential applications in electric devices[21]. The full range angular dependent MR at 16 T is shown in the inset of Fig. 4(a), where a two-fold rotation symmetric pattern can be clearly seen. Figure 4(b) shows the SdH oscillation spectra at different angles after subtracting the smooth background. The SdH oscillations can be observed in all angles, indicating an anisotropic 3D character of Fermi surface in $PtBi_2$. Figure 4(c) shows the angular dependencies of $F_\alpha$ and $F_\beta$. The frequencies of $\beta$ band involve a non-monotonic behavior as the field tilted from [111] direction to the applied current direction in the (111) plane, indicating the complex contour of the Fermi surface of this band. We here just consider the α-band by fitting the $F_\alpha$ with both 2D Fermi surface model and 3D standard ellipsoid Fermi surface model, and find that the 3D ellipsoid FS can reproduce the data better, consistent with expectation of anisotropic 3D FS.

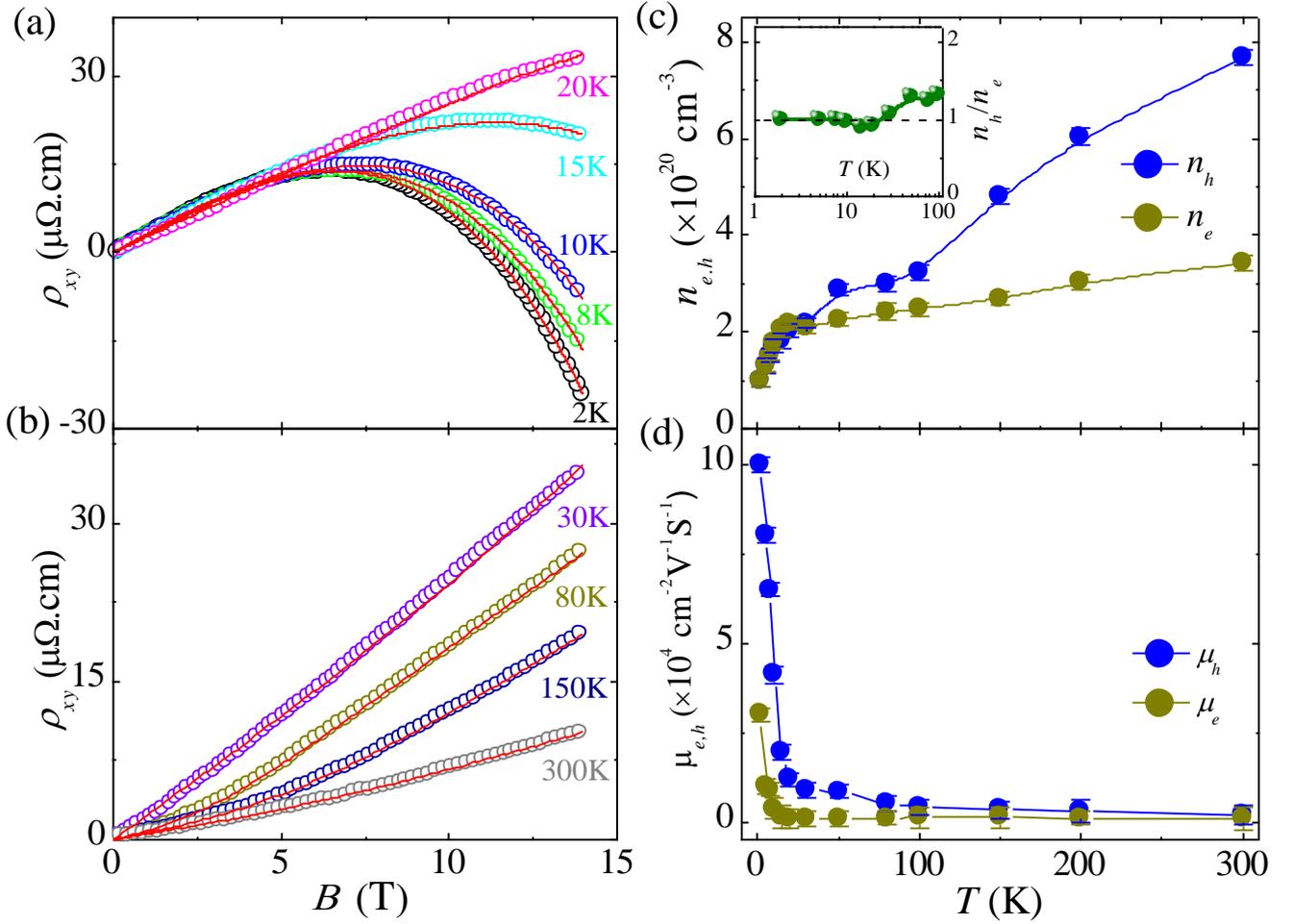

**FIG 5.** (a) and (b) Magnetic field dependence of Hall resistivity $\rho_{xy}$ at several typical temperatures. The red solid line are numerical fitting to the two-band model. **(c)** Temperature dependence of the carrier density $n_{e,h}$. Inset: the ratio of $n_h/n_e$ at low temperatures. (d) Temperature dependence of the mobility $\mu_{e,h}$.

To further investigate the transport properties of PtBi$_2$, we systemically measured the Hall resistivity at different temperatures. Several selected curves are shown in Fig. 5(a) and 5(b). Above 30 K, the Hall resistivity $\rho_{xy}$ is positive, indicating the hole-type carriers dominate the transport behavior. When lowering temperatures, the Hall resistivity keeps positive at low field but becomes negative in slope at higher fields, indicating a multiband effect in PtBi$_2$. For simplicity, here a typical two-band model was used to describe the Hall effect in the full temperature range,

$$\rho_{xy}(B) = \frac{B}{e} \frac{(n_h\mu_h^2 - n_e\mu_e^2) + (n_h - n_e)\mu_h^2\mu_e^2 B^2}{(n_h\mu_h + n_e\mu_e)^2 + (n_h - n_e)^2\mu_h^2\mu_e^2 B^2}, \tag{3}$$

where the $n_h$, $n_e$, $\mu_h$ and $\mu_e$ are the carrier concentrations and mobilities of electron and hole, respectively. The obtained $n_{h,e}$ and $\mu_{h,e}$ as a function of temperature are shown in Fig. 5(c) and 5(d). It can be found that both $n_h$ and $n_e$ decrease monotonically with decreasing temperature with a opposite trend of $\mu_{h,e}$. At 2 K, the carriers concentration reaches $n_h \approx 0.99(8) \times 10^{20} cm^{-3}$ and $n_e \approx 1.0(1) \times 10^{20} cm^{-3}$. Such densities are similar to the WTe$_2$,[14] LaSb[26], but two orders higher than the Cd$_3$As$_2$,[11] NbP,[21] and NbAs[35]. We note that, , the ratio of $n_h/n_e$ in PtBi$_2$ is also close to one below 30 K, indicating that the carriers of holes and electrons are nearly compensated in low temperature range which is similar to WTe$_2$[14] and NbAs[33]. We now turn to the classical transport mobility extracted from Hall resistivity. Above 30 K, the transport mobility for both carriers shows weak temperature dependence. However, as temperature decreases to 2 K, the mobility increases dramatically to $1.0 \times 10^5\ cm^2V^{-1}s^{-1}$ and $0.31 \times 10^5\ cm^2V^{-1}s^{-1}$ for hole and electron, respectively. This value is lower than NbP and Cd$_3$As$_2$, but comparable to WTe$_2$, LaSb, NbAs and TaAs. However, our MR ratio is higher than these materials and thus we inferred that the extremely large *RRR* (~1667) in our PtBi$_2$ may play an important role in boosting the highest XMR. Additionally, we note that the Hall mobility is about two orders higher than the quantum mobility obtained from the SdH oscillations. Such a deviation has been observed in other topological semimetals, such as Cd$_3$As$_2$[11] and NbAs[35], *et al.,* and might be associated with the different scattering processes. It is known that the quantum mobility is sensitive to all angle scattering processes while the classical transport mobility is only susceptible to large angle scattering process. The ratio $\tau_{tr}/\tau_Q$ is a measure of the relative importance of small angle scattering. According to the aforementioned experimental results, the transport lifetimes $\tau_{tr} = \frac{\mu m^*}{e}$ for the two pockets can be estimated to be $\tau_{tr}^\alpha \sim 3.64 \times 10^{-11} s$ and $\tau_{tr}^{\beta} \sim 1.1(5) \times 10^{-11} s$. The large ratio of $\tau_{tr}/\tau_Q > 100$, suggests that the backward scattering is reduced severely and the small angle scattering is dominated in PtBi$_2$.

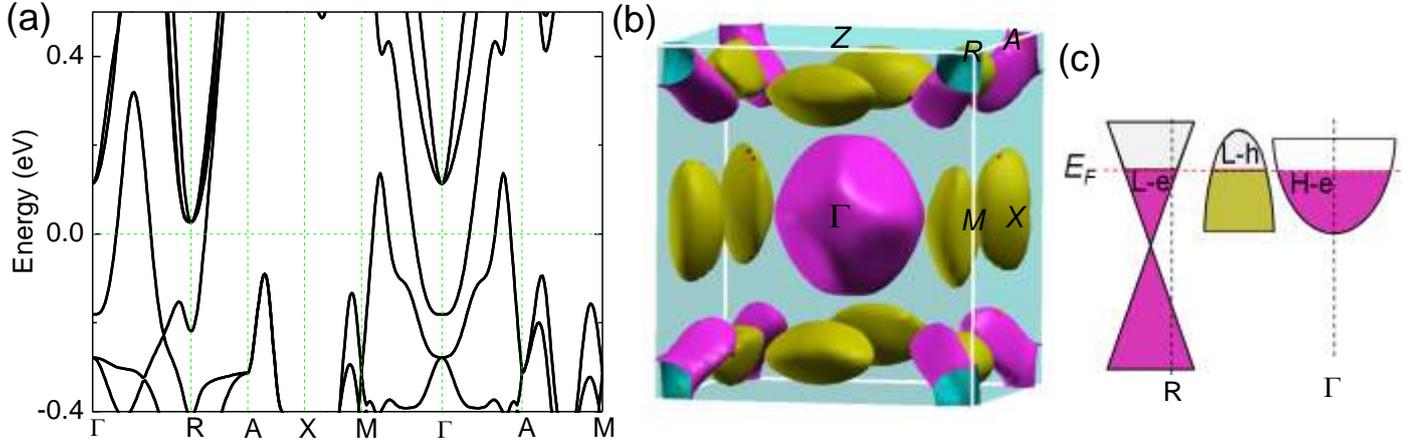

**FIG 6.** (a) and (b) The *ab initio* band structure and bulk Fermi surface of pyrite-type PtBi$_2$. (c) Schematic illustration of the semimetal hosted by pyrite-type PtBi$_2$. The pink and yellow colors corresponds to (b), and label the electron and hole bands, respectively. "L-e", "L-h" and "H-e" label the light electron, light hole and heavy electron with small and large effective mass, respectively.

To further understand the transport properties and verify the Dirac physics in PtBi$_2$, we have carefully performed *ab initio* calculation to obtain the electronic structure. The band structure is calculated by density function theory (DFT) and fully relaxed atomic coordinates. As the result shown in Fig. 6, there are three pockets, including two electron pockets locating around *Γ* point and *R* point respectively, and one hole pocket crossed by the *Γ-M* line (Actually, there are 12 hole pockets in the first Brillouin zone, but they are equivalent and can be counted one.). The coexistence of electron and hole is consistent with the Hall resistivity results. However, only two Fermi pockets have been identified from the FFT spectra of SdH oscillations (shown in Fig. 3). The reason might lie in the fact that the electrons from this pocket has much larger effective cyclotron mass, which is estimated to be 10 times of the effective mass of the hole band, *i.e.* 6.4 *m$_0$*, from the band structure, and respond insensitively to the external magnetic field, and thus cannot identified from the FFT spectra of SdH oscillations. From Fig. 6 (a), we find that there is a Dirac point located about ~230 meV below the Fermi level, which is close to our aforementioned experimental fitting result of $E_F^\beta$ ~ 292.4 meV. Meanwhile, the band top of the hole band to the Fermi level is about 130 meV, which is also close to $E_F^\alpha$ ~ 90.7 meV. This result also indicates the $\alpha$ and $\beta$ Fermi pockets correspond to the hole pocket and electron pocket around $R$ point, respectively. Such correspondence also coincides with the angular dependence of the oscillation frequency in Fig. 4 (c) that the β pocket is anisotropic while α pocket can be approximately described by ellipsoid. Finally, we plot a schematic diagram in Fig. 6 (c) to summarize all the key features of the Dirac semimetal of pyrite-type PtBi$_2$ with both experimental measurements and *ab initio* calculation. It is clear that there are two electron pockets from the linear semimetal band and normal quadratic band respectively, and a hole pocket from the normal quadratic band. This unique band structure is different to the well studied Cd$_3$As$_2$, WTe$_2$, TaAs and NbP, and should be related with the observation of extremely LMR in PtBi$_2$.

In conclusion, we have grown high-quality pyrite-type PtBi$_2$ single crystals and performed systemically quantum transport experiments under high magnetic fields. When beyond a critical magnetic field, the resistivity exhibits "turn-on" behavior and a plateau at low temperature range, which results in an extremely large MR of 11.2 million percent at $T =$ 1.8 K and $B = 33$ T. Such a large MR is even higher than WTe$_2$, LaSb and any other Dirac semimetals reported before. The crystals exhibit an ultrahigh mobility and clear Shubnikov-de Hass (SdH) quantum oscillations. Analysis of quantum oscillations and Hall resistivity provide detailed electronic structure information of PtBi$_2$, including the multiband characteristics, non-trivial Berry's phase that are consistent with our *ab initio* calculation, and provide evidence that $PtBi_2$ is topological semimetal candidate.

## Acknowledgments

The author thanks Professor Zhiqiang Mao, Dong Qian and Kun Yang for fruitful discussions. This work was supported by the Natural Science Foundation of China (Grant No.11174294, No.11574320, No.11374302, No.11204312, No.1147289, No.U1432251, No. 11674311 and No.11474030), and the National Key Research and Development Program of China No.2016YFA0401003, the 100 Talents Program of CAS, the program of Users with Excellence, the Hefei Science Center of CAS and the CAS/SAFEA international partnership program for creative research teams of China.

## References

[1] X. G. Wan, A. M.Turner, A. Vishwanath, and S. Y. Savrasov. *Phys. Rev. B* **83**, 205101 (2011).

[2] M. Orlita, , D. M. Basko, M. S. Zholudev, F. Teppe, W. Knap, V. I. Gavrilenko, N. N. Mikhailov, S. A. Dvoretskii, P. Neugebauer, C. Faugeras, C. Faugeras, A-L. Barra, G. Martinez and M. Potemski. *Nat. Phys.* **10**, 233-238 (2014).

[3] S. M. Young, S. Zaheer, J. C. Y. Teo, C. L. Kane, E. J. Mele, and A. M. Rappe. *Phys. Rev. Lett.* **108**, 140405 (2012).

[4] Z. J. Wang, Y. Sun, X. Q. Chen, C. Franchini, G. Xu, H. M. Weng, X. Dai, and Z. Fang. *Phys. Rev. B* **85**, 195320 (2012).

[5] Z. J. Wang, H. M. Weng, Q. S. Wu, X. Dai, and Z. Fang. *Phys. Rev. B* **88**, 125427 (2013).

[6] C. Fang, M. J. Gilbert, X. Dai, and B. A. Bernevig. *Phys. Rev. Lett.* **108**, 266802 (2012).

[7] Z. K. Liu, J. Jiang, B. Zhou, Z. J. Wang, Y. Zhang, H. M. Weng, D. Prabhakaran, S.-K. Mo, H. Peng, P. Dudin, T. Kim, Hoesch, Z. Fang, X. Dai, Z. X. Shen, D. L. Feng, Z. Hussain, and Y. L. Chen. *Nat. Mater.* **13**, 677–681 (2014).

[8] M. Neupane, S. Y. Su, R. Sankar, N. Alidoust, G. Bian, C. Liu, I. Belopolski, T. R. Chang, H. T. Jeng, H. Lin, A. Bansil, F. Chou, and M. Z. Hasan, *Nat. Commun.* **5**, 3786 (2014).

[9] S. Borisenko, Q. Gibson, D. Evtushinsky, V. Zabolotnyy, B. Buchner, and R. J. Cava. *Phys. Rev. Lett.* **113**, 027603 (2014).

[10] L. P. He, X. C. Hong, J. K. Dong, J. Pan, Z. Zhang, J. Zhang, and S. Y. Li. *Phys. Rev. Lett.* **113**, 246402 (2014).

[11] T. Liang, Q. Gibson, M. N. Ali, M. H. Liu, R. J. Cava, and N. P. Ong. *Nat. Mater.* **14**, 280–284 (2015).

[12] Z. K. Liu, B. Zhou, Y. Zhang, Z. J. Wang, H. M. Weng, D. Prabhakaran, S. -K. Mo, Z. X. Shen, Z. Fang, X. Dai, Z. Hussain, and Y. L. Chen. *Science* **343,** 864-867 (2014).

[13] S. Y. Xu, C. Liu, S. K. Kushwaha, R. Sanker, J. W. Krizan, I. Belopolski, M. Neupane, G. Bian, N. Alidoust, and T. R. Chang, H,-T, Jeng, C-Y. Huang, W-F. Tsai, H. Lin, P. P. Shibayev, F.-C. Chou, R. J. Cava, M. Z. Hasan. *Science* **347**, 294-298 (2015).

[14] M. N. Ali, J. Xiong, S. Flynn, J. Tao, Q. D. Gibson, L. M. Schoop, T. Liang, N. Haldolaarachchige, M. Hirschberger, N. P. Ong and R. J. Cava. *Nature* **514**, 205(2014).

[15] S. Y. Xu, I. Belopolski, N. Alidoust, M. Neupane, G. Bian, C. L. Zhang, R. Sankar, G. Q. Chang, Z. J. Yuan, C. C. Lee, S. M. Huang, H. Zheng, J. Ma, D. S. Sanchez, B. K. Wang, A. Bansil, F. C. Chou, P. P. Shibayev, H. Lin, S. Jia, and M. Z. Hasan. *Science* **349**, 613-617 (2015).

[16] B. Q. Lv, N. Xu, H. M. Weng, J. Z. Ma, P. Richard, X. C. Huang, L. X. Zhao, G. F. Chen, C. E. Matt, F. Bisti, V. N. Strocov, J. Mesot, Z. Fang, X. Dai, T. Qian, M. Shi, and H. Ding. *Nat. Phys.* **11**, 724-727 (2015).

[17] L. X. Yang, Z. K. Liu, Y. Sun, H. Peng, H. F. Yang, T. Zhang, B. Zhou, Y. Zhang, Y. F. Guo, M. Rahn, D. Prabhakaran, Z. Hussain, S. –K. Mo, X. Felser, B. Yan, and Y. L. Chen. *Nat. Phys.* **11**, 728-732(2015).

[18] S. Y. Xu, N. Alidoust, I. Belopolski, Z. J. Yuan, G. Bian, T.-R. Chang, H. Zheng, V. N. Strocov, D. S. Sanchez, G. Q. Chang, C. L. Zhang, D. X. Mou, Y. Wu, L. N. Huang, C. -C. Lee, S.-M. Huang, B. Wang, A. Bansil, H. Jeng, T. Neupert, A. Kaminski, H. Lin, S. Jia, and M. Z. Hasan. *Nat. Phys.* 11, 748-754 (2015).

[19] S. Y. Xu, I. Belopolski, D. S. Sanchez, C. Zhang, G. Chang, C. Guo, G. Bian, Z. Yuan, H. Lu, T. -R. Chang, P. P. Shibayev, M. L. Prokopovych, N. Alidoust, H. Zheng, C. -C. Lee, S. -M. Huang, R. Sankar, F. C. Chou, C. -H. Hsu, H. -T. Jeng, *et al*. *Sci. Adv.* **1**, e1501092 (2015).

[20] S. Souma, Z. W. Wang, H. Kotaka, T. Sato, K. Nakayama, Y. Tanaka, H. Kimizuka, T. Takahashi, K. Yamauchi, T. Oguchi, K. Segawa, Y. Ando. *Phys. Rev. B* **93**, 161112(R) (2016).

[21]C. Shekhar, A. K. Nayak, Y. Sun, M. Schmidt, M. Nicklas, I. Leermakers, U. Zeitler, Y. Skourski, J. Wosnitza, Z. K. Liu, Y. L. Chen, W. Schnelle, H. BOrrmannn, Y. Grin, C. Felser, and B. H. Yan. *Nat. Phys.* **11**, 645-649 (2015).

[22]A. Narayanan, M. D. Watson, S. F. Blake, N. Bruyant, L. Drigo, Y. L. Chen, D. Prabhakaran, B. Yan, C.

Felser, T. Kong, P. C. Canfield, and A. I. Coldea. *Phys. Rev. Lett.* **114**, 117201 (2015).

[23] Z. Wang, Y. Zheng, Z. X. Shen, Y. H. Lu, H. Y. Fang, F. Sheng, Y. Zhou, X. J. Yang, Y. P. Li, C. M. Feng, and Z. A. Xu. *Phys. Rev. B* 93, 121112(R) (2016).

[24] Q. D. Gibson, L. M. Schoop, L. Muechler, L. S. Xie, M. Hirschberger, N. P. Ong, R. Car, and R. J. Cava. *Phys. Rev. B* 91, 205128 (2015).

[25] Y. Y. Wang, Q. H. Yu, P. J. Guo, K. Liu, and T. L. Xia. *Phys. Rev. B* 94, 041103(R) (2016).

[26] F. F. Tafti, Q. D. Gibson, S. K. Kushwaha, N. Haldolaarachchige and R. J. Cava. *Nat. Phys.* 12, 272(2016).

[27] Y. F. Zhao, H. W. Liu, C. L. Zhang, H. C. Wang, J. F. Wang, Z. Q. Lin, Y. Xing, H. Lu, J. Liu, Y. Wang, S. M. Brombosz, Z. L. Xiao, S. Jia, X. C. Xie, and J. Wang. *Phys. Rev. X* 5, 031037 (2015).

[28] Y. L. Wang, L. R. Thoutam, Z. L. Xiao, J. Hu, S. Das, Z. Q. Mao, J. Wei, R. Divan, A. Luican-Mayer, G. W. Crabtree, and W. K. Kwok. *Phys. Rev. B* 92, 180402(R) (2015).

[29] E. Mun, H. Ko, G. J. Miller, G. D. Samolyuk, S. L. Bud'ko, and P. C. Canfield. *Phys. Rev. B* 85, 035135 (2012).

[30] H. Murakawa, M. S. Bahramy, M. Tokunaga, Y. Kohama, C. Bell, Y. Kaneko, N. Nagaosa, H. Y. Hwang, and Y. Tokura, *Science* 342, 1490 (2013).

[31] G. L. Zheng, J. W. Lu, X.D. Zhu, W. Ning, Y. Y. Han, H. W. Zhang, J.L. Zhang, C.Y. Xi, J.Y. Yang, H. F. Du, K. Yang, Y. H. Zhang, and M. L. Tian. *Phys. Rev. B* 93, 115414 (2016).

[32] D. Shoenberg, Magnetic Oscillations in Metals (Cambridge University Press, Cambridge, England, 1984).

[33] C. M. Wang, H. –Z. Lu, and S. –Q. Shen. *Phys. Rev. Lett.* 117, 077201 (2016).

[34] J. Xiong, Y. K. Luo, Y. H. Khoo, S. Jia, R. J. Cava, and N. P. Ong. *Phys. Rev. B* 86, 045314 (2012)

[35] Y. K. Luo, N. J. Ghimire, M. Wartenbe, H. Choi, M. Neupane, R. D. McDonald, E. D. Bauer, J. X Zhu, J. D. Thompson, and F. Ronning. *Phys. Rev. B* 92, 205134 (2015).